\documentclass[aps,preprint,showpacs,showkeys,preprintnumbers,eqsecnum,amsmath,amssymb]{revtex4}

\usepackage{dcolumn}
\usepackage[dvips]{epsfig}

\begin{document}
\title{\bf Collision of two general geodesic particles
around a Kerr-Newman black hole }

\author{
Changqing Liu$^{1,2}$, %\footnote{Electronic address: lcqliu2562@163.com}$,
Songbai Chen$^{1}$, %\footnote{Electronic address: csb3752@163.com}$,
and Jiliang  Jing\footnote{Corresponding author, Email:
jljing@hunnu.edu.cn}$^{1}$ %\footnote{Electronic address:jljing@hunnu.edu.cn}
}

\affiliation{1) Department of Physics, and Key Laboratory of Low Dimensional Quantum Structures \\ and Quantum Control of Ministry of Education, Hunan Normal
University, \\ Changsha, Hunan 410081, P. R. China}

\affiliation{2) Department of Physics and Information Engineering, \\
Hunan Institute of Humanities Science and Technology,\\ Loudi, Hunan
417000, P. R. China}

\begin{abstract}

We study collision of two general geodesic particles around the Kerr-Newman (KN) black
hole and get the center-of-mass (CM) energy of the non-marginally and marginally bound critical particles in the direct collision and LSO  collision  scenarios.
We find the constraint conditions that arbitrarily high CM energy can be obtained for the near-horizon collision of two general geodesic particles in the extremal KN black hole, and note that the charge  decreases the value of the latitude in which  arbitrarily high CM energy can occurs.  We also interpret why the high-velocity collision belt centers to the equator with the increase of the charge.

\end{abstract}


\maketitle

\newpage
\section{Introduction}
Banados, Silk and West (BSW)  \cite{BSW2009}  showed recently  that the extremal Kerr
black hole  surrounded by relic dark matter density spikes
could be regarded as a Planck-energy-scale collider, which
might allow us to explore  ultra high energy
collisions and astrophysical phenomena, such as the gamma ray bursts and the active
galactic nuclei. At the same time,  several authors
\cite{Berti_etal2009,Jacobson_Sotiriou2010,Thorne} pointed out that
the ultra-energetic collisions cannot occur near the
black hole in nature due to the astrophysical limitation.
For the kerr black hole, to circumvent the fine-tuning problem to obtain the CM
energy for the collision,  different scenarios, e.g.
multiple scattering \cite{ Grib_Pavlov2010_Kerr, Grib_Pavlov20102052}, the innermost stable circular orbit (ISCO) \cite{Harada} and the last stable orbit (LSO) \cite{Harada111,Sundararajan2008}, was proposed by several author.
In Refs. \cite{Zaslavskii2010_rotating,
Zaslavskii2010_charged,Zaslavskii3}, the author elucidated the
universal property of acceleration of particles for rotating black
holes and try to give a general explanation of this BSW mechanism.
The similar BSW mechanism  has also been found in other kinds of black holes and
special spacetime, such as stringy black hole \cite{Wei2}, Kerr-Newman black
holes \cite{WLGF2010} and Kaluza-Klein Black Hole \cite{Mao}, Kerr-Taub-NUT spacetime \cite{LIULCQ}, and naked singularity \cite{MandarPatil,MandarPatil1,MandarPatil2}.
The BSW mechanism also stimulated
some implications concerning the effects of gravity generated by
colliding particles \cite{Kimura}, the emergent flux from
particle collision near the Kerr black holes \cite{bsw1010}, and the numerical
estimation of the escaping
flux of massless particles created in collisions
around the Kerr black hole \cite{Williams}.

Recently, Harada \cite{Harada111} generalized
the analysis of the CM energy of two colliding particles
to general geodesic massive and massless particles in the Kerr black hole. They showed that, in the direct collision and LSO collision scenarios, the collision with an arbitrarily high CM energy can occur near the horizon of maximally rotating black holes not only at the equator but also on a belt centered at the equator. In this paper, we will extend Harada's work to the KN black hole. Besides the rotation parameter $a$, the
KN black hole has another parameter, the charge $q$. Thus, we will
demonstrate what effects of the charge on the CM energy for the particles in the near-horizon
collision of the general geodesic particles. And we also will give an interpretation why
the high-velocity collision belt centered at the equator with the increase of the charge.

This paper is organized as follows.
In Sec. II, we briefly review general geodesic particles
in the KN spacetime. In Sec. III, we obtain an expression for the CM energy
of two general geodesic particles at any spacetime point
and then obtain a general formula for the near-horizon collision.
We also discuss the collision with
an arbitrarily high CM energy in the direct collision and LSO collision scenarios, and
see what effects of the charge on the CM energy for the particles in the near-horizon
collision of the general geodesic particles. Sec. IV is
devoted to a brief summary. We use the units $c=G=1$ throughout the
paper.

\section{General geodesic orbits in Kerr-Newman spacetime}

The metric of the KN spacetime in the Boyer-Lindquist coordinates
can be expressed as
\begin{eqnarray}
 ds^{2}&=&\frac{\rho^{2}}{\Delta}dr^{2}+\rho^{2}d\theta^{2}
    +\frac{\sin^{2}\theta}{\rho^{2}}
    \Big[adt-(r^{2}+a^{2})d\phi\Big]^{2}\nonumber\\
    &&-\frac{\Delta}{\rho^{2}}\Big[dt-a \sin^{2}\theta d\phi\Big]^{2},
  \label{KNblackhole}
\end{eqnarray}
where
$
 \Delta=r^{2}-2r+a^{2}+q^{2},~~ ~~
 \rho^{2}=r^{2}+a^{2}\cos^{2}\theta,
$ and $M$, $a$ and $q$ are the mass, the rotation parameter and the
electric charge. The event horizon of the KN black hole is given by
$r_{H} =M + \sqrt{M^2-a^2-q^2}$, and the extremal case corresponds
to the condition $a^2+q^2 =M^2$. The angular velocity of the KN
black hole is
\begin{equation}
\Omega_{H}=\frac{a}{(r_H^2+a^2)}=\frac{a}{(2M^2-q^2+2M\sqrt{M^2-q^2-a^2})}.
\end{equation}
The nonvanishing
contravariant components $g^{\mu\nu}$ of the metric are
\begin{eqnarray}
g^{tt}&=&-\frac{(r^{2}+a^{2})^{2}-a^{2}\Delta \sin^{2}\theta}{\rho^{2}\Delta},
\quad g^{t\phi}=g^{\phi t}=-\frac{2M ar-q^2}{\rho^{2}\Delta}, \nonumber \\
g^{rr}&=& \frac{\Delta}{\rho^{2}}, \quad
g^{\theta\theta}=\frac{1}{\rho^{2}}, \quad
g^{\phi\phi}= \frac{\Delta-a^{2}\sin^{2}\theta}{\Delta \rho^{2}\sin^{2}\theta}.
\label{eq:inverse_metric}
\end{eqnarray}

The general geodesic motion of
massive particles in the KN spacetime
was analyzed in Refs. \cite{Carter1968,MTW1973}. So
we here  briefly review general geodesic particles in the KN spacetime.
The Hamiltonian for the geodesic motion is given by
\begin{equation*}
{\cal H}[x^{\alpha},p_{\beta}]=\frac{1}{2}
g^{\mu\nu}p_{\mu}p_{\nu},
\end{equation*}
where $p_{\mu}$ is the conjugate momentum to $x^{\mu}$.
Let $S=S(\lambda,x^{\alpha})$ be the action as a function of
the parameter $\lambda$ and coordinates $x^{\alpha}$, the conjugate momentum $p_{\alpha}$ is
described by $p_{\alpha}=\frac{\partial S}{\partial x^{\alpha}}$.

Then the corresponding Hamilton-Jacobi equation is
\begin{equation}\label{carter191}
-\frac{\partial S}{\partial \lambda}={\cal H}\left[x^{\alpha},\frac{\partial S}{\partial x^{\beta}}\right]=\frac{1}{2}g^{\mu\nu}\frac{\partial S}{\partial x^{\mu}}\frac{\partial S}{\partial x^{\nu}}.
\end{equation}
For the KN black hole, the action can be expressed as
\begin{equation}
S=\frac{1}{2}m ^{2}\lambda-Et+L\phi+\sigma_{r}\int^{r}dr\frac{\sqrt{R}}{\Delta}+\sigma_{\theta}\int^{\theta}d\theta\sqrt{\Theta},
\label{eq:variable_separation}
\end{equation}
where the constants $m$, $E$ and $L$ are the rest mass,
energy and angular momentum of the particle. The sign functions $\sigma_{r}=1(-1)$ and $\sigma_{\theta}=1(-1)$ correspond to the outgoing (ingoing) geodesics.
From Eq. (\ref{carter191}) we can obtain \cite{Carter1968}
\begin{eqnarray}
\rho^{2}\frac{dt}{d\lambda}&=&-a(aE\sin^{2}\theta-L)+\frac{(r^{2}+a^{2})P}{\Delta},
\label{eq:dtdlambda}\\
\rho^{2}\frac{dr}{d\lambda}
&=&\sigma_{r} \sqrt{R}, \label{eq:drdlambda}\\
\rho^{2}\frac{d\theta}{d\lambda}&=& \sigma_{\theta}\sqrt{\Theta},
\label{eq:dthetadlambda}\\
\rho^{2}\frac{d\phi}{d\lambda}&=& -\left(aE-\frac{L}{\sin^{2}\theta}\right)+\frac{aP}{\Delta}.
\end{eqnarray}
with
\begin{eqnarray}
\Theta&=&\Theta(\theta)={\cal Q}-\cos^{2}\theta\left[a^{2}(m ^{2}-E^{2})+\frac{L^{2}}{\sin^{2}\theta}\right],
\label{eq:Theta}\nonumber \\
R&=& R(r)=P(r)^{2}-\Delta(r) [m ^{2}r^{2}+(L-aE)^{2}+{\cal Q}],
\label{eq:R}\nonumber \\
P&=&P(r)= (r^{2}+a^{2})E-aL. \label{eq:Theta_R_P}
\end{eqnarray}
where $\cal{Q}$ is the Carter constant \cite{Carter1968}. Then the radial equation
for the timelike particle moving along geodesics is
\begin{equation}
\frac{1}{2}u^ru^r+V_{\rm eff}(r)=0, \label{eq:eom}
\end{equation}
where the effective potential is defined by
\begin{eqnarray}
V_{\rm eff}(r)=-\frac{R(r)^2}{2\rho ^4}
\label{eq:effective_potential}.
\end{eqnarray}
The circular orbit of the particle can be found by the conditions
\begin{eqnarray}\label{veercondition}
V_{\rm eff}(r)=0,~~~~ \frac{dV_{\rm eff}(r)}{dr}=0.
\end{eqnarray}
Because we are interested in causal geodesics, we also
need to impose the condition $\frac{dt}{d\lambda}>0$.
As $r\rightarrow r_{H}$ for the timelike particle, this condition reduces to
\begin{equation*}
E\ge\frac{aL}{2(a^2+r_H^2)}= \Omega_{H}L,
\end{equation*}
which shows us that the angular momentum
must be equal to or smaller than the critical value $L_{c}\equiv \Omega_{H}^{-1}E$.

\section{CM energy of two colliding general geodesic particles in KN spacetime}

In this section, we will study the CM energy for the collision of
two particles moving along  general geodesic in the KN spacetime and the high-velocity collision belts on the extremal KN black hole.

\subsection{\label{sec:CM_energy_11}
CM energy of two colliding particles  in KN spacetime}

Let us now consider two uncharged colliding particles with rest masses $m_1$
and $m_2$.  We assume that two particles 1 and 2 are located at the same
spacetime point with the four momenta $p_{(i)}^{a}=m_{(i)}u^{a}_{(i)}$.
The CM energy $E_{\rm cm}$ of the two particles is shown by \cite{Harada111}
\begin{eqnarray}
E_{\rm cm}^{2}=m_{1}^{2}+m_{2}^{2}
-2g^{ab}p_{(1)a}p_{(2)b}.\label{eq:center-of-mass_energy_different_mass}
\end{eqnarray}
With the help of Eqs. ({\ref{eq:inverse_metric}) and~(\ref{eq:center-of-mass_energy_different_mass}), the
CM energy of two colliding general geodesic particles in the KN spacetime is
\begin{eqnarray}
E_{\rm cm}^{2}&=&m_{1}^{2}+m_{2}^{2}+\frac{2}{\rho^{2}}\left[\frac{P_{1}P_{2}-
\sigma_{1r}\sqrt{R_{1}}\sigma_{2r}\sqrt{R_{2}}}{\Delta}-
\frac{(L_{1} -a\sin^{2}\theta E_{1})(L_{2}-a\sin^{2}\theta E_{2})}{\sin^{2}\theta}\right. \nonumber \\
&& \left. -\sigma_{1\theta}\sqrt{\Theta_{1}}\sigma_{2\theta}\sqrt{\Theta_{2}}\right].
\label{eq:E_cm_explicit}
\end{eqnarray}

Now we will investigate the properties of the CM energy as the
radius $r$ approaches to the horizon $r_H$ of the non extremal black
hole. $\sigma_{1r}$ and $\sigma_{2r}$ have the same sign on the horizon $r=r_{H}$.
Note that both denominator and the numerator of the fraction $\frac{P_{1}P_{2}-
\sqrt{R_1}\sqrt{R_2}}{\Delta}$ on the right-hand side of
Eq.~(\ref{eq:E_cm_explicit}) vanishes at $r_H$. Using l'Hospital's rule
and taking into account $r_H^2-2r_H-a^2-Q^2=0$,  the CM energy of two general
geodesic particles in the near-horizon limit can be expressed as
\begin{eqnarray}
E_{\rm cm}^{2}(r_H)&=&m_{1}^{2}+m_{2}^{2}+
\frac{1}{\rho^2_{r_H}}
\left[(m_{1}^{2}r_{H}^{2}+{\cal K}_{1})\frac{E_{2}-\Omega_{H}L_{2}}{E_{1}-\Omega_{H}L_{1}}+(m_{2}^{2}r_{H}^{2}+{\cal K}_{2})\frac{E_{1}-\Omega_{H}L_{1}}{E_{2}-\Omega_{H}L_{2}} \right. \nonumber \\
&& \left. -\frac{2(L_{1}-a\sin^{2}\theta E_{1})(L_{2}-a\sin^{2}\theta E_{2})}{\sin^{2}\theta}
-2\sigma_{1\theta}\sqrt{\Theta_{1}}\sigma_{2\theta}\sqrt{\Theta_{2}}\right],
\label{eq:general_formula}
\end{eqnarray}
 where ${\cal K}_{i}={\cal Q}_{i}+(L_i-aE_i)^2$ for the particles $i$.
 We can now find that the necessary condition
to obtain an arbitrarily high CM energy is that $E_i-\Omega_H L_i=0$, i.e.,
either of the two particles must possess the critical  angular momentum $L_{ic}=\frac{E_i}{\Omega_H}$.

 For the colliding particles with the same rest mass $m_0$
moving on the equatorial plane,
 Eq.~(\ref{eq:general_formula}) reduces to
\begin{eqnarray}
E_{\rm cm}^{2}(r_H)&=&2m_{0}^{2}+\frac{1}{r_{H}^{2}}
\left\{\left[m_{0}^{2}r_{H}^{2}+(L_{1}-aE_{1})^{2}\right]\frac{E_{2}-\Omega_{H}L_{2}}{E_{1}-\Omega_{H}L_{1}}\right.\nonumber \\
&& \left. +\left[m_{0}^{2}r_{H}^{2}+(L_{2}-aE_{2})^{2}\right]\frac{E_{1}-\Omega_{H}L_{1}}{E_{2}-\Omega_{H}L_{2}}-2(L_{1}-a E_{1})(L_{2}-a E_{2})\right\}.
\label{eq:Harada_Kimura_variant}
\end{eqnarray}
If we further assume that the colliding particles have the same energy $E_1=E_2$,
 Eq.~(\ref{eq:Harada_Kimura_variant}) becomes
\begin{equation*}
\frac{E_{\rm cm}(r_H)}{2m_{0}}=\sqrt{1+\frac{(L_1-L_2)^2}{(L_1-L_c)(L_2-L_c)}\frac{L_c}{4a}}.
\end{equation*}
 which coincides with the result in Ref. \cite{WLGF2010}.

\subsection{\label{subsec:belt}
The high-velocity collision belts in extremal KN black hole}
We are now in the position to study
the collision of two particles with an arbitrarily high CM
energy. In Ref. \cite{Harada111},
T. Harada. and M. Kimura  divide the collision scenario into four types according to
the effective potentials for the critical particles:  The first type is the
direct collision with the conditions $R'(r_{H})=0,~R(r_{H})=0$, and $R''(r_{H})>0 $;
the second one is LSO \cite{Sundararajan2008} collision with the conditions
$R'(r_{H})=0,~ R(r_{H})=0$, and $R''(r_{H}=0 $; the third one is
multiple scattering with the conditions $R'(r_{H})=0,~R(r_{H})=0$, and $R''(r_{H})<0 $;
and the fourth one is also multiple scattering but with the condition that
$R'(r_{H})=0$ and $R'(r_{H})<0 $. Here we take Harada-Kimura's classification and concentrate our attention on the
direct collision and LSO collision scenarios.

For the critical particles defined by its angular momentum
$L_c=\frac{E}{\Omega _H}$, from  Eq.~(\ref{eq:R}), we can easily
find that the condition $R(r_{H})=0$ is satisfied by both the
nonextremal and extremal black holes. However, for the first
derivative, $R'(r_{H})=-2(r_H-1)(2m^2r^2_H+{\cal K})$, the condition
$R'(r_{H})=0$ holds only for the extremal KN spacetime, and the
condition $R'(r_{H})<0$ is true for the nonextremal KN black hole
which shows that direct collision and LSO collision scenarios for
the critical particles in the nonextremal KN black hole do not
exist. Therefore, we only consider the critical particles in the
extremal KN black hole. Using  Eq.~(\ref{eq:R}), $R''(r_H)$ becomes
 \begin{equation}
R''(r_{H})=2\left[\left((4-\frac{M^2}{a^2})E^{2}-m^{2}\right)M^{2}-{\cal Q}\right].
\end{equation}
Then, $R''(r_{H})\geq0$ shows that
\begin{eqnarray}
\left((4-\frac{M^2}{a^2})E^{2}-m^{2}\right)M^{2}\geq {\cal Q}.
\end{eqnarray}
From Eq. (\ref{eq:Theta_R_P}), we find the following condition
\begin{equation}
\cos^{2}\theta\left[a^{2}(m ^{2}-E^{2})+\frac{(M^{2}+a^2)^2E^{2}}{a^2\sin^{2}\theta}\right]
\le  {\cal Q} \le \left((4-\frac{M^2}{a^2})E^{2}-m^{2}\right)M^{2},
\label{eq:range_of_Q}
\end{equation}
where $L_{c}=\frac{(M^2+a^2)E}{a}$ was used.
Using Eq.~(\ref{eq:range_of_Q}) and taking $M^2=q^2+a^2$ for the
extremal black hole, the following condition must be satisfied
\begin{equation}
(m ^{2}-E^{2})\sin^{4}\theta+\left((7E^{2}-2m^2)(a^2+q^2)
+a^2E^2\right)\sin^{2}\theta-\frac{(q^2+2a^2)^2E^{2}}{a^2}\ge 0.
\label{eq:inequality}
\end{equation}
For the marginally bound orbit $m ^{2}= E^{2}$, we can  find
\begin{equation}\label{sdfsdfd}
\sin\theta\ge \sqrt{\frac{(q^2+2a^2)^2}{a^2(6a^2+5q^2)}}.
\end{equation}
The result confirms that charge $q$ of the black
hole indeed influences on the angle of
the collision of two general   critical particles.
We obtain $sin\theta \geq \sqrt{\frac{2}{3}}$ when $q=0$, which
coincides with the kerr case \cite{Harada111}.
If we set $M=1$, the right hand of the inequality (\ref{sdfsdfd})
 becomes $ \frac{(2-q^2)^2}{(1-q^2)(6-q^2)}$,
 which  monotonically increases with the charge $q$. Therefore we can the
get the maximum  charge $q$
when $\sin\theta =1$, i.e., $q=\sqrt{\frac{2}{3}}$ ( the corresponding $a$ is
$\frac{1}{\sqrt3}$).
From the Fig. \ref{fg:latitude}, we find that the highest latitude ($\frac{\pi}{2}-\theta $) decreases
with the increase of the charge $q$.

\begin{figure}
\includegraphics[width=0.7\textwidth]{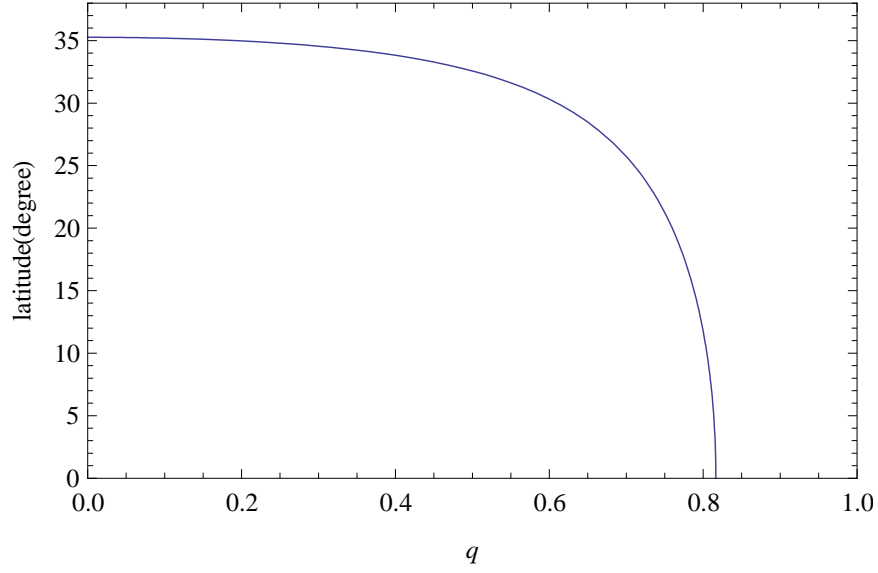}
\caption{\label{fg:latitude}
the variation of The highest value of
the latitude for the critical  particles  with charge $q$ of the
extremal KN black hole
in the marginally bound orbit. Here we set $M=1$ .
}
\end{figure}

For the bound ($m^2-E^2>0$) and the unbound ($m^2-E^2<0$) particles,
we can find that $\theta$ must satisfy
\begin{equation}
\sin\theta\ge \sqrt{\frac{-B+\sqrt{B^2-4AC}}
{2A}},
\label{eq:belt_general}
\end{equation}
where $ A=m^2-E^2, ~~B=(7E^{2}-2m^2)(a^2+q^2)+a^2E^2, ~~
C=-\frac{(q^2+2a^2)^2E^{2}}{a^2}. $ The highest absolute value of
the latitude is shown in Fig. \ref{fg:latitude112} as a function of
the specific energy of the particle. We find that the charge $q$
decreases the highest value of the latitude in which the arbitrarily
high CM energy can occurs.
\begin{figure}
\includegraphics[width=0.7\textwidth]{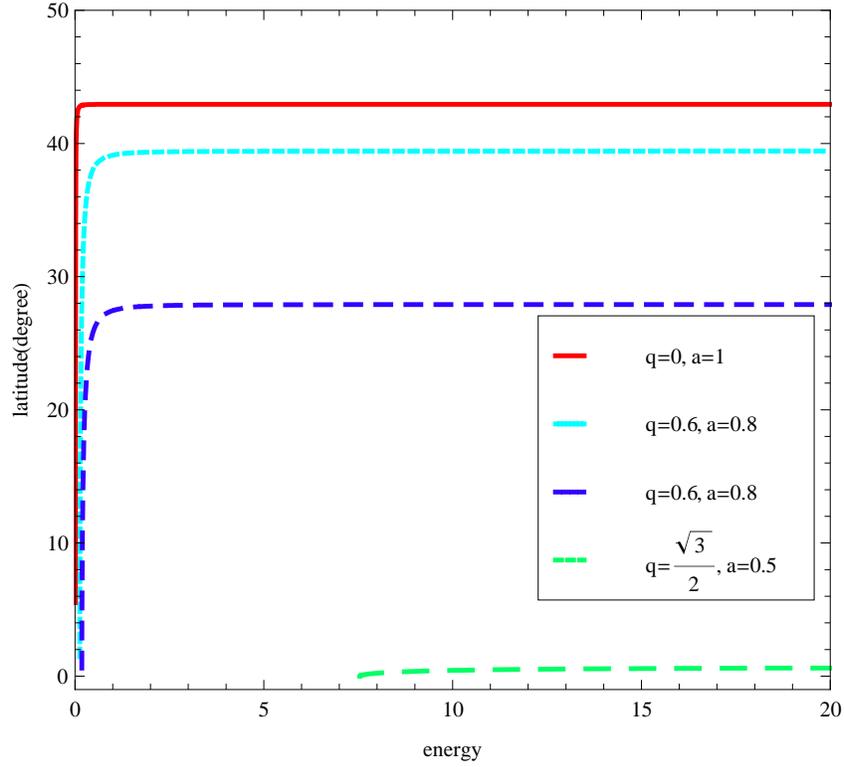}
\caption{\label{fg:latitude112}
The variation of the highest value of
the latitude for the critical  particles  with the energy of the particles
in the non marginally bound orbit with different $q$ and
$a$. Here we set $M=1$ and $m=0.2$.
}
\end{figure}
Thus, the highest value of the latitude $\alpha$ of
both the bound  and unbound critical particles is given as
\begin{equation}
\alpha(E,m)=\mbox{arccos}\sqrt{\frac{-B+\sqrt{B^2-4AC}}{2A}}.
\end{equation}
In the limit $E\to \infty$,  Eq.~(\ref{eq:belt_general}) gives
\begin{equation}\label{pppppd}
\sin\theta\ge \sqrt{\frac{(8a^2+7q^2)-
\sqrt{\frac{a^2(8a^2+7q^2)^2-4(2a^2+q^2)^2}{a^2}}}{2}} .
\end{equation}
Then we can get the constraint on the value of  the rotation
parameter $a$ and charge $q$:  $ 1\geq a\geq\frac{1}{2}, ~~0\leq
q\leq\frac{\sqrt{3}}{2}. $ For a massless particle with critical
angular momentum, we can also get the same result. In Fig.
\ref{fg:lataa} we  also find that  with the increase of the  charge
$q$ the highest value of the latitude for the critical massless
particles decreases.

When the charge $q\rightarrow 0$, the inequality  (\ref{pppppd})
reduces to the result in Ref. \cite{Harada111}
\begin{equation}
\sin\theta\ge \sqrt{3}-1.
\end{equation}

\begin{figure}
\includegraphics[width=0.7\textwidth]{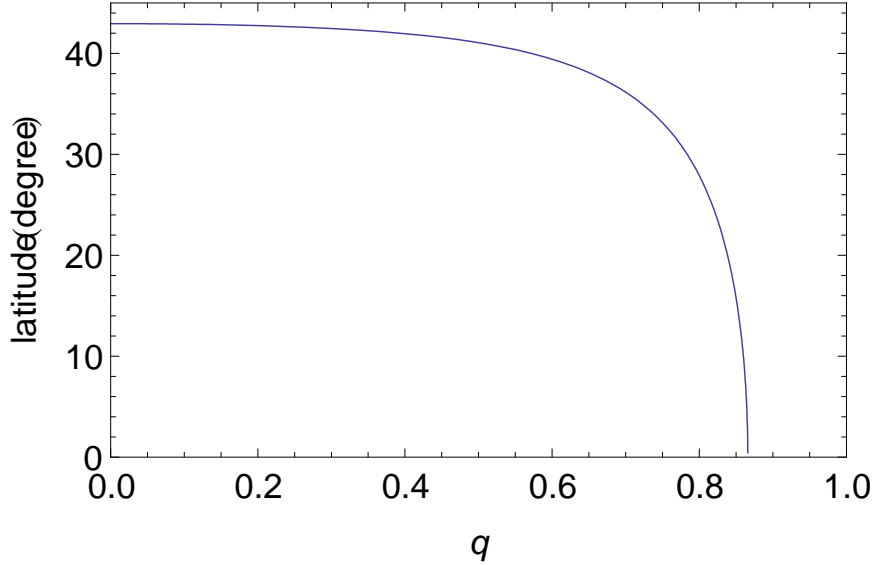}
\caption{\label{fg:lataa}
The variation of the highest value of
the latitude for the critical massless particles with charge $q$ of the
extremal KN black hole
in the  non marginally bound orbit. Here we take $M=1$.
}
\end{figure}

It is interesting to note that the constraint  conditions for the
charge $q$ and rotation parameter $a$ are  $0\leq q\leq
\sqrt{\frac{2}{3}}, ~ 1\geq a\geq \sqrt{\frac{1}{3}}$ for marginally
bound particles, which are different from $0\leq q\leq
\frac{\sqrt{3}}{2}, ~ 1\geq a\geq \frac{1}{2}$ for non-marginally
bound critical particles.

\subsection{\label{subsec:interpretation}
Interpretation of constrains to value of $a$ and
$q$ in extremal KN black hole} Now we try to give an
interpretation of the constrains on the value of rotation parameter
$a$ and the charge $q$.  According to the Penrose
mechanism, in general, the BSW mechanism allows rotational energy of
a rotating black hole to be extracted by scattered particles
escaping from the ergosphere to infinity, and the ergosphere will
become thinner as the energy extracted from the black hole. Thus, the thickness of the ergosphere  plays an important role in the process of obtaining high CM energy. The fact shows us that the constrains on  parameters $a$ and $q$ to obtain arbitrarily high CM
energy  corresponds to the constrains on the thickness of the ergosphere.

The infinite redshift surface of the KN black hole
is
\begin{equation*}
r^\infty_{\pm}=M\pm\sqrt{M^2-q^2-a^2cos^2\theta}.
\end{equation*}
The ergosphere is the region bounded by the event horizon $r_H$ and the outer stationary limit surface $r^\infty_+$. The thickness of the ergosphere for the extremal KN black hole is
\begin{equation}
H \equiv r^\infty_+-r_H=a sin\theta.
\end{equation}
Using Eqs. (\ref{eq:inequality}) and (\ref{eq:belt_general}), we can get the minimum thickness of the ergosphere for the marginally bound and massless particle with the critical angular momentum to obtain arbitrary high CM energy. The minimum thickness of the ergosphere is shown in Fig. \ref{pppe}.

\begin{figure}[ht]
\begin{center}
\includegraphics[width=8cm]{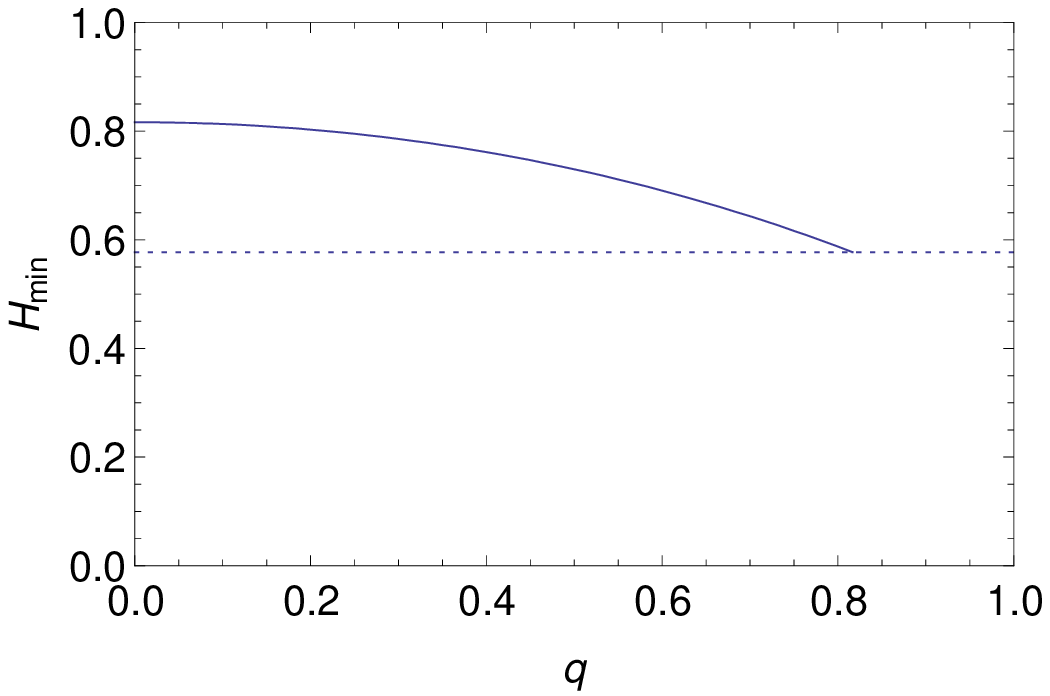}
\includegraphics[width=8cm]{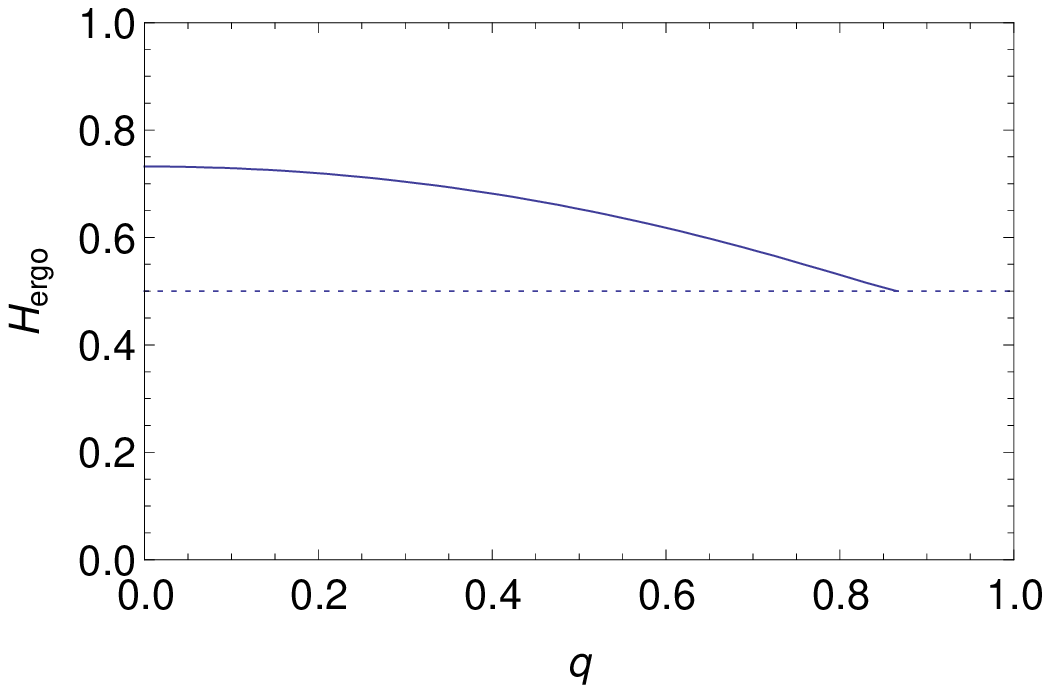}
\caption{\label{pppe}The variation of the minimum thickness of the ergosphere $H_{min}$  with charge $q$ for the marginally bound critical  particle (left) and massless critical particle (right) for the extremal KN black hole.}
\end{center}
\end{figure}

From the Fig. 4 we find that the minimum thickness of the ergosphere satisfies $H_{min}\geq \frac{1}{\sqrt{3}}$ for  marginally bound particles. However,
it satisfies $H_{min}\geq \frac{1}{2}$ for the massless particles.

\begin{figure}
\includegraphics[width=0.7\textwidth]{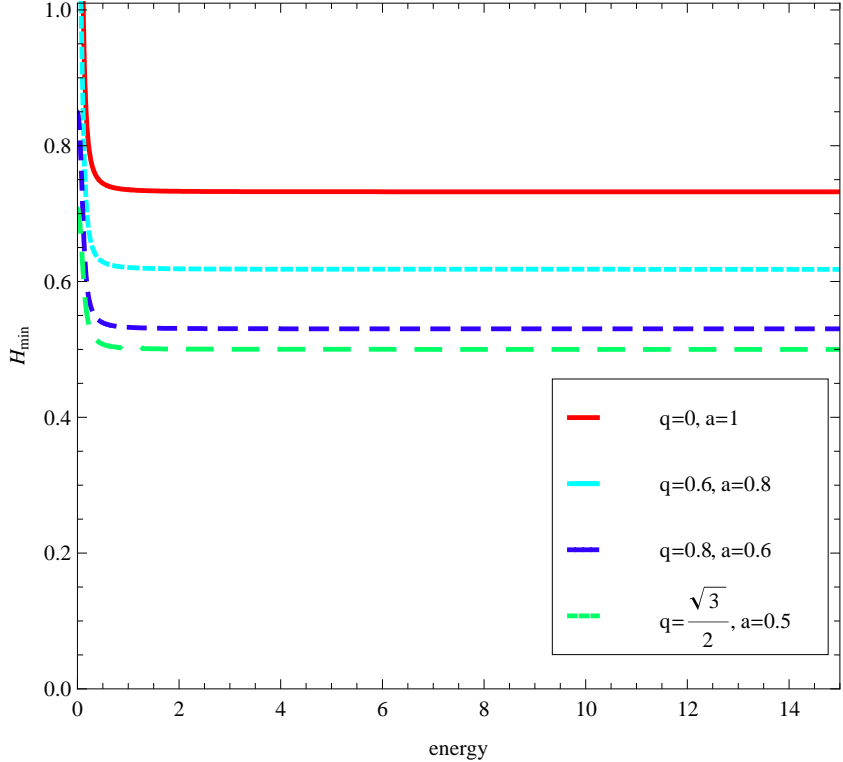}
\caption{\label{fg:lat1}
the variation of the minimum thickness of the ergosphere
 with the specific energy vs the charge $q$ and
rotation parameters $a$ of the
extremal KN black hole
in the  non marginally bound orbit, we set $M=1,m=0.2$ .
}
\end{figure}

For the non marginally bound particle, the variation of $H_{min}$ as the function of
specific energy  with different charge $q$ is shown in Fig. 5, which shows that
the minimum thickness of the ergosphere decreases with increase of the charge $q$.

In fact, when the extremal KN black hole carries more the charge $q$, the rotation energy is more less. The collision with an arbitrarily high CM energy must have enough rotation energy. Thus, the high-velocity collision belt centers at the equator with the increase of the  charge $q$ because the equator has the maximum thickness of the ergosphere  $H_{max}=a$ and the rotation energy. When the charge $q\rightarrow \frac{\sqrt{3}}{2}$ for the non marginally bound particle, the high-velocity collision only occurs at the equator where the maximum thickness of the ergosphere $H_{max}=a$ which coincides with
the minimum thickness $H_{min}$.  Above interpretation is also true for the marginally bound particle. Therefore, we argue that the constrains on the value of
rotation parameter $a$ and the charge $q$ in order to obtain  arbitrarily high CM energy is corresponding to the constrains on the thickness of the ergosphere.

\section{summary}

We studied collision of two general geodesic particles and get the formula for the CM energy of the non-marginally and marginally bound critical particles in the direct collision and LSO  collision  scenarios. Our study showed that arbitrarily high CM energy can be obtained for the near-horizon collision of two general geodesic particles in the extremal KN black hole  under the two conditions: (1) either of the impingement particles has the critical angular momentum $L_c=\frac{E}{\Omega_H}$; and (2) the charge $q$ and rotation parameter $a$ satisfy  $0\leq q\leq \sqrt{\frac{2}{3}}, ~ 1\geq a\geq \sqrt{\frac{1}{3}}$ for marginally bound particles,  and $0\leq q\leq \frac{\sqrt{3}}{2}, ~ 1\geq a\geq \frac{1}{2}$ for non-marginally bound critical particles.
We find that the presence of the charge $q$  will decrease the value of the latitude in which  arbitrarily high CM energy can occurs.  Finally, we present an interpretation why the high-velocity collision belt centers to the equator with the increase of the charge $q$, i.e, the constrains on the value of $a$ and $q$ in order to obtain  arbitrarily high CM energy is corresponding to the constrain on the thickness of the ergosphere.

\begin{acknowledgments}

This work was supported by the National Natural Science Foundation
of China under Grant No 10875040;  the key project of the
National Natural Science Foundation of China under Grant No
10935013;  the National Basic Research of China under Grant No.
2010CB833004;   PCSIRT under Grant No.  IRT0964;
and the Construct Program  of the National Key Discipline.

\end{acknowledgments}

\end{document}